\documentstyle[11pt,newpasp,twoside]{article}
\def\edcomment#1{\iffalse\marginpar{\raggedright\sl#1\/}\else\relax\fi}
\reversemarginpar

\begin{document}
\title{Searching for spiral shocks} 

\author{L. Morales-Rueda, T. R. Marsh} 

\affil{Dept. of Physics and Astronomy, University of Southampton,
  Highfield, Southampton, SO17 1BJ, UK}

\begin{abstract}
  Spiral shocks in cataclysmic variables (CVs) are the result of tidal
  interactions of the mass donor star with the accretion disc. Their
  study is fundamental for our understanding of angular momentum
  transfer in discs. In our quest to learn how widespread amongst
  binaries spiral shocks are, and how their presence depends upon
  orbital period and mass ratio (as they are created by direct
  interaction with the donor star), we have obtained spectra of a
  large sample of CVs during their high-mass-transfer states. We find
  that 24 out of the 63 systems observed are candidates for containing
  spiral shocks. 5 out of those 24 CVs have been confirmed as showing
  shocks in the disc during outburst.
\end{abstract}

\section{Introduction}
In 1996, Sawada, Matsuda, \& Hachisu suggested that the angular
momentum in accretion discs could be transferred by means of shocks,
tidally excited by the donor star. These shocks were seen for the
first time in IP~Peg during outburst by Steeghs, Harlaftis, \& Horne
(1997). The shocks have since been seen in several other outbursts of
IP~Peg (Harlaftis et al. 1999; Morales-Rueda, Marsh, \& Billington
2000). The spiral pattern is not seen during quiescence (Marsh \&
Horne 1990), consistent with the idea that a hot disc is needed to
lower the Mach number.

Spiral shocks have been successfully modelled using numerical
calculations (Steeghs \& Stehle 1999) but there are still some
unanswered questions about these spiral patterns: How do they evolve
with time? Do they form every outburst?  Are they still present on the
return to quiescence? Above all we would like to know how widespread
amongst other systems these shocks are, and how they depend upon the
system's orbital period and mass ratio.

At present, IP~Peg provides the best example of these shocks, with
some signs in the systems SS~Cyg (Steeghs et al.\ 1996), EX~Dra
(Joergens, Spruit, \& Rutten 2000), U~Gem (Groot 2001) and WZ~Sge
(Steeghs et al. this volume). Two common features seen in the spectra
of these 5 systems during outburst are that they contain emission
lines, and the shocks are stronger in He{\sc ii}\,4686\AA. The small
number of systems where shocks have been seen is not surprising as
there are remarkably few extensive datasets of outburst spectra.  This
work aims at solving this problem by obtaining spectra of dwarf novae
during outburst and identifying those with strong emission lines,
mainly He{\sc ii}\,4686\AA.

\section{Observations}

We obtained spectra of 63 dwarf novae in outburst. The spectra can be
seen in Morales-Rueda \& Marsh (2001). Table 1 gives a list of the
objects observed. Systems that show He{\sc ii}\,4686\AA\ in emission
are marked with a star.

\begin{table}
\caption{Sample of dwarf novae observed during outburst.}
\begin{tabular}{llllll}
\tableline
FO~Aql   & HL~CMa *  & V1504~Cyg & AY~Lyr    & GK~Per *  & QW~Ser  \\
VZ~Aqr   & SV~CMi    & AB~Dra    & CY~Lyr    & KT~Per *  & UZ~Ser  \\
FO~And * & AM~Cas    & EX~Dra *  & V419~Lyr  & TZ~Per    & DI~UMa  \\
LX~And   & GX~Cas    & IX~Dra    & V493~Lyr  & TY~Psc *  & ER~UMa *\\
RX~And   & KU~Cas    & AH~Eri    & V426~Oph *& VZ~Pyx *  & IY~UMa *\\
SS~Aur * & TU~Crt *  & IR~Gem    & BI~Ori *  & AW~Sge    & SS~UMa  \\
CR~Boo   & EM~Cyg *  & U~Gem *   & CN~Ori    & RZ~Sge    & SS~UMi *\\
AT~Cnc   & SS~Cyg *  & AH~Her    & V1159~Ori & WZ~Sge *  & VW~Vul *\\
AY~Cnc   & V516~Cyg  & V589~Her  & HX~Peg *  & V893~Sco *& \\
YX~Cnc   & V542~Cyg *& X~Leo     & IP~Peg *  & RY~Ser *  & \\
Z~Cam    & V792~Cyg  & RZ~LMi    & RU~Peg    & NY~Ser    & \\
\tableline
\tableline

\end{tabular}
\end{table}

\section{Results}
24 out of 63 dwarf novae observed show He{\sc ii}\,4686\AA\ in
emission.  5 of these 24 systems have been seen to show spiral shocks
in their accretion discs during outburst. These spiral shocks are most
prominent in He{\sc ii}\,4686\AA.

We find a correlation between the inclination of the systems and the
strength of their emission lines.  We do not find any clear
correlation between the masses of the components of the system and the
equivalent width of He{\sc ii}, but should keep in mind that the
uncertainties in the mass measurements are very large. We do not find
a correlation between orbital period and the presence of He{\sc
  ii}\,4686\AA. See Morales-Rueda \& Marsh (2001) for further details.

\end{document}